\documentclass[ip]{jjap3}

\def\gsim{\mathop {\vtop {\ialign {##\crcr 
$\hfil \displaystyle {>}\hfil $\crcr \noalign {\kern1pt \nointerlineskip } 
$\,\sim$ \crcr \noalign {\kern1pt}}}}\limits}
\def\lsim{\mathop {\vtop {\ialign {##\crcr 
$\hfil \displaystyle {<}\hfil $\crcr \noalign {\kern1pt \nointerlineskip } 
$\,\,\sim$ \crcr \noalign {\kern1pt}}}}\limits}

\title{New Quantum Criticality Revealed under Pressure}
\author{Shinji Watanabe and Kazumasa Miyake$^{1}$}
\inst{Department of Basic Sciences, Kyushu Institute of Technology, Kitakyushu, Fukuoka 804-8550, Japan \\
$^{1}$Toyota Physical and Chemical Research Institute, Nagakute, Aichi 480-1192, Japan}
\abst{Unconventional quantum critical phenomena observed in Yb-based periodic crystals such as YbRh$_2$Si$_2$ and $\beta$-YbAlB$_4$ have been one of the central issues in strongly correlated electron systems. 
The common criticality has been discovered in quasicrystal Yb$_{15}$Au$_{51}$Al$_{34}$, which 
surprisingly persists under pressure at least up to $P=1.5$~GPa. 
The $T/H$ scaling where the magnetic susceptibility can be expressed as a single scaling function of the ratio of temperature $T$ and magnetic field $H$ has been discovered in the quasicrystal, which is essentially the same as that observed in $\beta$-YbAlB$_4$. 
Recently, the $T/H$ scaling as well as the common criticality has been also observed even in the approximant crystal Yb$_{14}$Au$_{51}$Al$_{35}$ under pressure. 
The theory of critical Yb-valence fluctuation gives a natural explanation for these striking phenomena in a unified way. 
}

\begin{document}
\maketitle

\section{Introduction}
Quantum critical phenomena in itinerant-electron systems have attracted much attention in condensed matter physics. 
By changing control parameter of the system, continuous transition temperature of the ordered state can be suppressed to 0 K, at which a quantum critical point (QCP) is realized. Pressure is one of the most important control parameters, which can change the electronic state to tune the system to the QCP. 
Near the QCP, non Fermi-liquid behavior emerges in a series of physical quantities such as resistivity $\rho(T)$, specific heat $C(T)$, magnetic susceptibility $\chi(T)$, and NMR relaxation rate $(T_1T)^{-1}$, which are referred to quantum critical phenomena. Quantum criticality emerging near the magnetic QCP in itinerant electron systems has been well understood from the spin-fluctuation theory developed by Moriya~\cite{Moriya,MK,MT}, which has been endorsed by the renormalization-group theory by Hertz~\cite{Hertz} and Millis~\cite{Millis} (see Table~\ref{tb:QCP}).

\begin{table}
\caption{Quantum criticality of conventional and unconventional types. 3d AF denotes conventional three-dimensional antiferromagnetic criticality. Unconventional criticality observed in Yb-based periodic crystals and quasicrystal Yb$_{15}$Au$_{51}$Al$_{34}$. CVF denotes quantum criticality of critical valence fluctuations (CVF) emerging for temperature region $T>T_{0}$ with $T_{0}$ being characteristic temperature of CVF. $\zeta$ takes the value for $0.5\le\zeta\le 0.7$ depending on material and temperature range (see Ref.~\citen{WM2010} for details).} 
\label{tb:QCP}
\begin{tabular}{llllll}
\Hline
Theories and materials & $\rho(T)$ & $C(T)/T$ & $\chi(T)$ & $1/(T_1T)$ & Refs. \\
\Hline
3d AF & $T^{3/2}$ & const.$-T^{1/2}$ & const.$-T^{1/4}$ & $T^{-3/4}$ & \citen{Moriya,MT,Hatatani} \\ 
YbCu$_{3.5}$Al$_{1.5}$ & $T^{1.5}\to T$ & $-\log{T}$ & $T^{-0.66}$ & * & \citen{Bauer1997,Bauer2000} \\
YbRh$_2$Si$_2$ & $T$ & $-\log{T}$ & $T^{-0.6}$ & $T^{-0.5}$ & \citen{Trovarelli,Ishida} \\
$\beta$-YbAlB$_4$ & $T^{1.5}\to T$ & $-\log{T}$ & $T^{-0.5}$ & * & \citen{Nakatsuji,Matsumoto} \\
Yb$_{15}$Au$_{51}$Al$_{34}$ & $T$ & $-\log{T}$  & $T^{-0.51}$ & $T^{-0.51}$ & \citen{Deguchi,Watanuki}\\
CVF & $T$ & $-\log{T}$ & $T^{-\zeta}$ & $T^{-\zeta}$ & \citen{WM2010,WM2012} \\
\Hline
\end{tabular}
\end{table}

On the other hand, unconventional quantum criticality has been observed in Yb-based metal YbCu$_{5-x}$Al$_{x}$ near $x=1.5$~\cite{Bauer1997,Bauer2000} (see Table~\ref{tb:QCP}) where sharp Yb-valence crossover has been observed~\cite{Bauer1997}. The similar unconventional criticality has been also observed in the heavy-electron metals YbRh$_2$Si$_2$~\cite{Trovarelli,Ishida} and $\beta$-YbAlB$_4$~\cite{Nakatsuji,Matsumoto}, as listed in Table~\ref{tb:QCP}. 
As a possible origin of the unconventional criticality, the theory of critical Yb-valence fluctuation has been proposed~\cite{WM2010, WM2012}, which gives a unified explanation for the measured quantum critical phenomena (see Table~\ref{tb:QCP}). 

Recently, the common unconventional quantum critical phenomena have been discovered in the Yb-based quasicrystal Yb$_{15}$Au$_{51}$Al$_{34}$ at ambient pressure $P=0$ and zero-magnetic field $H=0$~\cite{Deguchi,Watanuki} (see Table~\ref{tb:QCP}). 
Surprisingly, the quantum criticality persists even under pressure at least up to $P=1.5$~GPa~\cite{Deguchi}. 
Recently, the common quantum criticality has been observed in the approximant crystal Yb$_{14}$Au$_{51}$Al$_{35}$ when pressure is tuned~\cite{Matsukawa2016}.  
Interestingly, a new type of scaling called ``$T/H$ scaling" has been discovered in the pressurized approximant crystal and the 
quasicrystal~\cite{note_QC_TB}, where the magnetic susceptibility is expressed as a single scaling function of the ratio of temperature $T$ and magnetic field $H$~\cite{Matsukawa2016}.
This behavior is essentially the same as that observed in the periodic crystal $\beta$-YbAlB$_4$~\cite{Matsumoto}. Furthermore, the $T/H$ scaling behavior has been also observed even in the approximant crystal when pressure is tuned~\cite{Matsukawa2016}. 
The theory of critical valence fluctuation (CVF) has been shown to give a natural explanation for these striking phenomena in a unified way. 

This paper reviews 
a new type of quantum criticality arising from the CVF, mainly focusing on new aspects of quantum critical phenomena revealed under pressure. 
In section~2, the theory of CVF will be explained. Experimental indication of valence instability will be discussed in section~3. The paper will be summarized in section~4.

\section{Theory of critical valence fluctuation (CVF)}

As the simplest minimal model for the Ce- and Yb-based heavy-electron systems, 
we consider the extended periodic Anderson model
\begin{eqnarray}
{\cal H}_{\rm EPAM}&=&\sum_{{\bf k}\sigma}\varepsilon_{\bf k}
c_{{\bf k}\sigma}^{\dagger}c_{{\bf k}\sigma}
+\varepsilon_{\rm f}\sum_{i\sigma}n_{i\sigma}^{\rm f}
+\sum_{{\bf k}\sigma}\left(V_{\bf k}
f_{{\bf k}\sigma}^{\dagger}c_{{\bf k}\sigma}
+{\rm h.c.}
\right)
+U\sum_{i}n_{i\uparrow}^{\rm f}n_{i\downarrow}^{\rm f}
\nonumber
\\
&+&U_{\rm fc}\sum_{i\sigma\sigma'}n_{i\sigma}^{\rm f}n_{i\sigma'}^{\rm c}
\label{eq:EPAM}
\end{eqnarray}
with 
$n_{i\sigma}^{\rm f}=f_{i\sigma}^{\dagger}f_{i\sigma}$ and $n_{i\sigma}^{\rm c}=c_{i\sigma}^{\dagger}c_{i\sigma}$. 
The first line is the so-called periodic Anderson model. 
The conduction electron with the energy band $\varepsilon_{\bf k}$ and f electron with the f level $\varepsilon_{\rm f}$ hybridizes with the hybridization $V_{\bf k}$. 
Onsite Coulomb repulsion between f electrons is denoted by $U$ in the 4th term. 
The second line indicates the inter-orbital Coulomb repulsion between f and conduction electrons $U_{\rm fc}$, which plays an important role for causing the first-order valence transition.
For example, in Ce metal which exhibits the $\gamma$-$\alpha$ transition, there are 4f and 5d bands at the Fermi level~\cite{Pickett}. Since both the 4f and 5d orbitals are located at the same Ce site, 
$U_{\rm fc}$ has a considerable value and cannot be neglected (see Ref.~\citen{WTMF2009} for details). 

\begin{figure}
\begin{center}
\includegraphics[width=15cm]{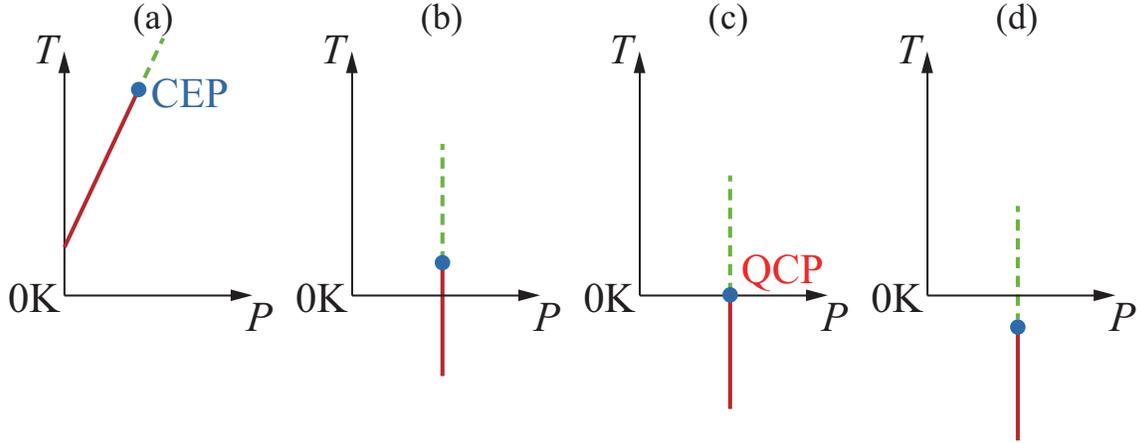}
\end{center}
\caption{Schematic phase diagrams of valence transition and crossover in the temperature $T$ and pressure $P$ phase diagram. The first-order valence transition (solid line) terminates at the critical end point (CEP) (solid circle), from which the valence-crossover line (dashed line) extends. 
As $U_{\rm fc}$ in the extended periodic Anderson model, Eq.~(\ref{eq:EPAM}), decreases, the critical temperature decreases, corresponding to the cases from (a) to (d). 
In (c), the quantum CEP i.e., quantum critical point (QCP) is realized.}
\label{fig:T_P}
\end{figure}

Actually, theoretical calculations on the basis of ${\cal H}_{\rm EPAM}$, Eq.~(\ref{eq:EPAM}), have shown that the first-order 
valence transition takes place for large enough $U_{\rm fc}$~\cite{OM2000_B,WIM2006,Saiga,WM_CeRhIn5,Kojima} and the corresponding $T$-$P$ phase diagram is illustrated in  Fig.~\ref{fig:T_P}(a)~\cite{WM2011}. 
In Fig.~\ref{fig:T_P}(a), the first-order valence transition terminates at the critical end point (CEP), from which the valence-crossover line extends. 
As $U_{\rm fc}$ decreases, the critical temperature decreases, which corresponds to the cases from Fig.~\ref{fig:T_P}(a) to Fig.~\ref{fig:T_P}(d). 
When the critical temperature is located at $T=0$~K, the quantum CEP is realized, which is called the quantum critical point (QCP) of the valence transition (see Fig.~\ref{fig:T_P}(c)). 
At the QCP, the CVF diverges, which can cause instability of the electronic state. Indeed, superconducting correlation has been shown to be enhanced in the Kondo regime near the QCP by theoretical calculations in ${\cal H}_{\rm EPAM}$~\cite{OM2000,WIM2006}. 
The CVF-mediated superconductivity is favorably compared with the pressure-induced superconductivity in CeCu$_2$Si$_2$~\cite{Jaccard}, CeCu$_2$Ge$_2$~\cite{Holms}, and CeCu$_2$(Si$_{1-x}$Ge$_{x}$)$_2$~\cite{Yuan} (see Ref.~\citen{Miyake2007} for details).

\subsection{Mode-mode coupling theory of critical valence fluctuations}

To clarify the quantum critical phenomena of the CVF arising from the QCP of the valence transition,  
we have constructed the mode-mode coupling theory of CVF taking account of local correlation effects of f electrons due to the on-site Coulomb repulsion $U$, starting from ${\cal H}_{\rm EPAM}$, Eq.~(\ref{eq:EPAM})~\cite{WM2010}. We have found that the almost dispersionless CVF mode emerges near ${\bf q} ={\bf 0}$ in the momentum space. 
This causes extremely small coefficient $A$ in the valence susceptibility, 
\begin{eqnarray}
\chi_{\rm v}(q,i\omega_{m})^{-1}\approx\eta+Aq^2+C\frac{|\omega_{m}|}{q}, 
\label{eq:chiv}
\end{eqnarray}
giving rise to 
extremely small characteristic temperature of CVF, $T_0$, 
which is defined by 
\begin{eqnarray}
T_{0}\equiv\frac{Aq_{\rm B}^3}{2{\pi}C}
\label{eq:T0}
\end{eqnarray}
with $q_{\rm B}$ being the wave number of the Brillouin zone. 
Then, even at temperatures low enough compared to the effective (renormalized) Fermi temperature $T_{\rm F}^{*}$, 
the temperature region scaled by $T_0$, $T/T_0>1$, can be regarded as ``high-temperature". Therefore, the region of temperature, $T_{\rm F}^{*}\gg T\gsim T_{0}$, can be regarded as ``high-temperature"  region, where 
new quantum criticality emerges in the physical quantities such as the valence susceptibility  $\chi_{\rm v}$, specific-heat coefficient $C(T)/T$, resistivity $\rho(T)$, and NMR relaxation rate $(T_{1}T)^{-1}$~\cite{WM2010,WM2012}.
In Table~\ref{tb:QCP}, quantum valence criticality, which appears in the temperature region $T\gsim T_{0}$, is summarized. 
Near the valence QCP, the magnetic susceptibility $\chi$ and the valence susceptibility $\chi_{\rm v}(0,0)$ are enhanced by the common many-body effect caused by the inter-orbital Coulomb repulsion $U_{\rm fc}$ so that both show the same criticality~\cite{WM2008,WM2010}. Then, huge Wilson ratio appears. 
It is noted that in the region $T\lsim T_{0}$, electronic resistivity behaves as $\rho(T)\sim T^{1.5}$ in dirty systems and $\rho(T)\sim T^{5/3}$ in clean systems~\cite{WM2010}. 
This gives a unified explanation for the unconventional quantum criticality observed in Yb-based metals, as shown in Table~\ref{tb:QCP}~\cite{WM2012,MW2014}.

\subsection{Robust quantum criticality in quasicrystal Yb$_{15}$Au$_{51}$Al$_{34}$ under pressure}

\begin{figure}
\begin{center}
\includegraphics[width=15cm]{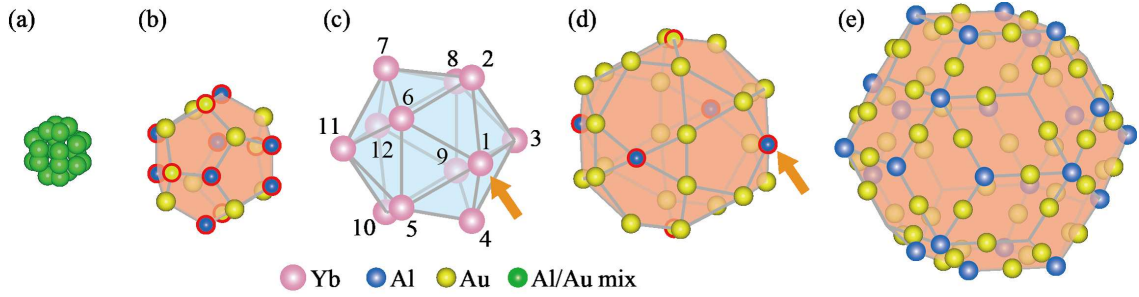}
\end{center}
\caption{(Color online) Concentric shell structures of Tsai-type cluster, which is a core structure of 
quasicrystal and approximant crystal. (a) 1st shell, (b) 2nd shell, (c) 3rd shell, (d) 4th shell, 
and (e) 5th shell. In (b) and (d), Al/Au mixed sites are framed in red. In (c), the number indicates the $i$-th Yb site.}
\label{fig:YbAuAl}
\end{figure}

To get insight into the mechanism of the robust quantum criticality in Yb$_{15}$Au$_{51}$Al$_{34}$  under pressure, 
let us  start with analyzing the lattice structure of the quasicrystal. 
The quasicrystal is constituted of the Yb-Au-Al cluster which has concentric shell structures from the 1st to 5th shells, as shown in Figs.~\ref{fig:YbAuAl}(a)-(e), respectively.  
In the 3rd shell, there are 12 Yb atoms. In the 1st, 2nd, and 4th shells, Al/Au mixed sites exist. Namely, Al or Au atom is located with existence ratio of 
Al/Au being 7.8~$\%$/8.9~$\%$ (1st shell), 62~$\%$/32~$\%$ (2nd shell), and 59~$\%$/41~$\%$ (4th shell)~\cite{Ishimasa}. 
In Figs.~\ref{fig:YbAuAl}(b) and \ref{fig:YbAuAl}(d), the Al/Au mixed sites framed in red are illustrated as a representative case following the existence ratio.
There also exists approximant crystal which has periodic arrangement of the body centered cubic (bcc) structure of the Yb-Au-Al cluster with concentric shell structures shown in Figs.~\ref{fig:YbAuAl}(a)-\ref{fig:YbAuAl}(e)~\cite{Ishimasa}. 

As discussed in section~2.1, locality of the CVF mode is considered to be the key origin of emergence of the new type of quantum criticality. Namely, charge transfer between the Yb site and surrounding atoms is considered to play a key role, which is essentially local. Hence, let us concentrate on the Yb-Au-Al cluster.

Recent measurement performed by replacing Al with Ga in the quasicrystal has revealed that quantum critical behavior in physical quantities disappears~\cite{Matsukawa2014}. This suggests that conduction electrons at Al sites contribute to the quantum critical state. 
Hence, we consider the extended Anderson model with f orbital at Yb site and conduction orbital at Al site on the Yb-Au-Al culster as the simplest minimal model~\cite{WM2013}. 
This model has essentially the same structure as the extended periodic Anderson model with the $U_{\rm fc}$ term ${\cal H}_{\rm EPAM}$, Eq.~(\ref{eq:EPAM}), which exhibits the quantum valence criticality.

\begin{figure}
\begin{center}
\includegraphics[width=15cm]{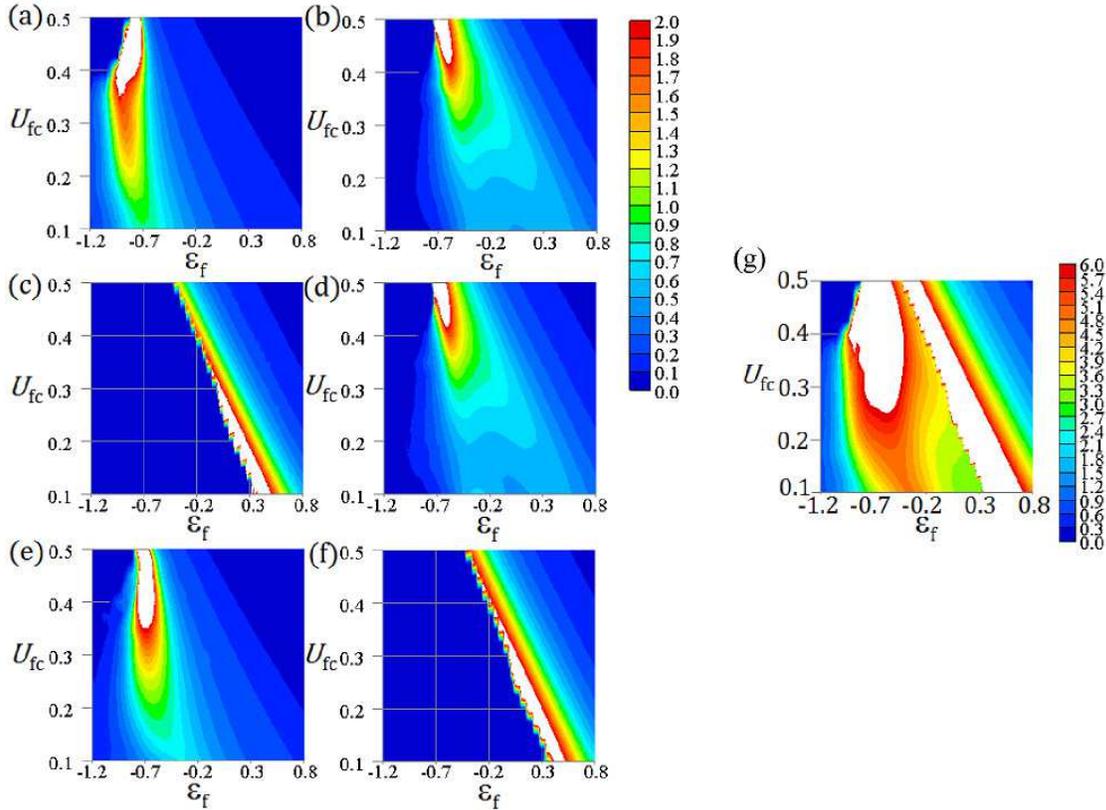}
\end{center}
\caption{(Color online) Contour plot of valence susceptibility $\chi_{{\rm v}i}$ for (a) $i=1$, 
(b) $i=2$, (c) $i=3$, (d) $i=4$, (e) $i=5$, and (f) $i=6$ calculated for the Yb-Au-Al cluster~\cite{WM2013}. In each white region, CVF diverges at the QCP of valence transition. (g) Total valence susceptibility $\chi_{\rm v}=\sum_{i=1}^{12}\chi_{{\rm v}i}$.}
\label{fig:chiv_i}
\end{figure}

By using the slave-boson mean-field theory for $U=\infty$, we determined the ground-state phase diagram of the Yb-Au-Al cluster for a typical parameter set of heavy electrons~\cite{WM2013}. 
The result of the contour plot of the valence susceptibility for each f site, which is indicated as the $i$-th site in Fig.~\ref{fig:YbAuAl}(c),  
$
\chi_{{\rm v}i}=-\frac{\partial n_{i}^{\rm f}}{\partial\varepsilon_{\rm f}}
$
is shown in Figs.~\ref{fig:chiv_i}(a)-\ref{fig:chiv_i}(f), respectively, representing  CVF in the phase diagram of the f-level $\varepsilon_{\rm f}$ and the inter-orbital Coulomb repulsion $U_{\rm fc}$.
We see that the valence QCPs appear as spots, which are located inside of the white islands. 
This is because the strength of the f-c hybridization at each f site is different due to the Al/Au mixed sites. 
The contour plot of the total valence susceptibility $\chi_{\rm v}=\sum_{i=1}^{12}\chi_{{\rm v}i}$ 
is shown in Fig.~\ref{fig:chiv_i}(g).  
An important result is that the valence QCPs appear as spots in the phase diagram, whose critical regions are overlapped to be unified, giving rise to a wide quantum critical region. 

When pressure is applied to Yb compounds, the f level $\varepsilon_{\rm f}$ in the hole picture decreases and $U_{\rm fc}$ increases in general. Hence, in case that applying pressure follows the line located in the enhanced CVF region, robust quantum criticality appears under pressure. 
Emergence of a wide critical region also gives a natural explanation for why quantum criticality appears without tuning control parameters in this material~\cite{WM2013}.

\subsection{The same criticality in approximant crystal by pressure tuning}

In section~2.2, we analyzed the fundamental nature of the Yb-Au-Al cluster which is the core structure of both the quasicrystal and approximant crystal. 
Essentially local nature of the CVF offers an interesting possibility that the same criticality appears in the approximant crystal when pressure is tuned~\cite{WM2013}.  

By further considering outer concentric shells to the Yb-Au-Al cluster shown in Fig.~\ref{fig:YbAuAl}, the quasicrystal is constructed, where the valence QCPs are expected to appear as widespread and condensed spots like, for instance, the Andromeda Galaxy. 
On the other hand, in the approximant crystal, Yb 12 cluster is periodically arranged, giving rise to 24 valence QCP spots in the bulk limit at most, since 24 Yb atoms are located inside the bcc unit cell. 
Hence, quantum critical region in the quasicrystal is expected to be vaster than that in the approximant crystal~\cite{WM2015}.

Recently, these theoretical predictions have been actually observed experimentally~\cite{Matsukawa2016}. When pressure is applied to the approximant crystal Yb$_{14}$Au$_{51}$Al$_{35}$
and is tuned to $P=1.96$~GPa, it has been discovered that the magnetic susceptibility 
exhibits the quantum critical behavior $\chi\sim T^{-0.5}$, which is the same as that  
in the quasicrystal Yb$_{15}$Au$_{51}$Al$_{34}$. 
It has been also observed that the quantum critical region in the $T$-$P$ phase diagram is much wider in the quasicrystal~\cite{Deguchi} than that in the approximant crystal~\cite{Matsukawa2016}. 

Emergence of the common quantum criticality in the quasicrystal and approximant crystal 
suggests that the key origin arises from the Yb-Au-Al cluster with strong locality of the CVF 
as underlying mechanism because essentially ``local" CVF does not depend on the details of lattice structures.
Namely, emergence of the quantum valence criticality itself does not depend on periodic crystal or quasicrystal while extent of the critical regions in their phase diagrams can be different as noted above.  
The significance of the Yb-Au-Al cluster is also reinforced by the experimental fact that 
the simple Kondo-disorder scenario due to the Al/Au mixed sites is incompatible with the robustness of the critical exponent under pressure~\cite{Deguchi}. 
The appearance of the common criticality even in the pressurized approximant crystal 
also offers serious difficulties in the Kondo-disorder scenario for the quasi periodicity~\cite{Andrade,Otsuki}. 
It remains to be resolved whether the quantum critical behavior of physical quantities such as  $\chi(T)$, $C(T)/T$, $\rho(T)$, and $(T_{1}T)^{-1}$ listed in Table~\ref{tb:QCP} can be explained by the ``Kondo disorder scenario" in a unified way. 

In the approximant crystal, the magnetically-ordered phase has been observed for $P\gsim P_{\rm c}=2$~GPa~\cite{Matsukawa2016}. However, this magnetic order is considered to be secondary effect on the unconventional criticality since there exists no magnetic order anywhere of the $T$-$P$ phase diagram 
from $P=0$ at least up to $P=1.5$~GPa in the quasicrystal~\cite{Deguchi}. 
Theoretically, it has been shown that the valence-crossover pressure $P_{\rm v}^{*}$, which is defined as the pressure at the dashed line in Figs.~\ref{fig:T_P}(c) and \ref{fig:T_P}(d), coincides with magnetic transition pressure $P_{\rm c}$ in ${\cal H}_{\rm EPAM}$ in the case of 
rather small f-c hybridization strength in Eq.~(\ref{eq:EPAM})~\cite{WM_CeRhIn5,WM2011}. 
The $T$-$P$ phase diagram of the approximant crystal is in accord with this result.

\subsection{Valence crossover ``region" in Yb- and Ce-based quasicrystal}

\begin{figure}
\begin{center}
\includegraphics[width=10cm]{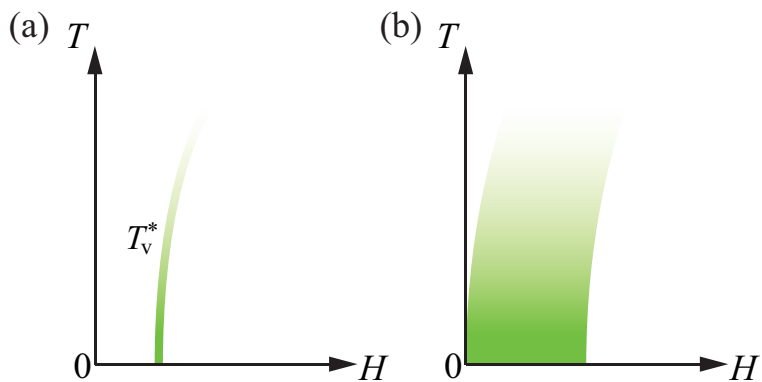}
\end{center}
\caption{(Color online) (a) Valence-crossover line arising from the CEP of the first-order valence transition at negative-$T$ side (see Fig.~\ref{fig:T_P}(d)) in periodic crystal. (b) Valence-crossover ``region" arising from the condensed CEP's at negative-$T$ side in quasicrystal.}
\label{fig:Tv_H}
\end{figure}

When the magnetic field is applied to ${\cal H}_{\rm EPAM}$, Eq.~(\ref{eq:EPAM}), the valence QCP starts to shift to the smaller-$U_{\rm fc}$ direction in the ground-state phase diagram of the $\varepsilon_{\rm f}$-$U_{\rm fc}$ plane~\cite{WM2008,WTMF2009}. Hence, in case the system is located in the Kondo regime close to the valence QCP for $U<U_{\rm QCP}$, a field-induced valence-crossover line is expected to appear, as shown in Fig.~\ref{fig:Tv_H}(a). 
As $H$ varies across $T_{\rm v}^{*}$, crossover from the Kondo regime with relatively-larger $n_{\rm f} (\lsim 1)$ to the mixed-valence regime with relatively-smaller $n_{\rm f}$ occurs. 
This valence-crossover effect arising from the CEP occurs in addition to the ordinary Zeeman effect which makes $n_{\rm f}$ increase monotonically to reach the $n_{\rm f}=1$ state to earn the Zeeman energy. Hence, negative contribution to the monotonic increase in $n_{\rm f}$ is expected to be remarkable at the $T_{\rm v}^{*}$ line in Fig.~\ref{fig:Tv_H}(a), especially 
for lower temperatures, since the effect of finite temperature easily makes the valence crossover be smeared out.  

On the other hand, in the quasicrystal Yb$_{15}$Au$_{51}$Al$_{34}$, 
the valence QCPs are densely condensed, giving rise to a valence-crossover ``region" 
with a certain width, as shown in Fig.~\ref{fig:Tv_H}(b)~\cite{WM2015}. 
In this case, the valence-crossover induced negative contribution to the monotonic increase in $n_{\rm f}$ by the Zeeman effect occurs for a certain $H$ region. 
It should be noted here that 
absolute values of the $n_{\rm f}$ changes, caused by the valence crossover and the Zeeman effect, respectively, depend on the details of each material.  
Hence, at least, some counter effect to the monotonic increase in $n_{\rm f}$ ordinary seen as the Zeeman effect is expected to be conspicuous for a certain-$H$ ``region", especially for low temperatures. 

\subsection{$T/H$ scaling in pressurized approximant crystal and quasicrystal}

To clarify the origin of quantum critical phenomena commonly observed in the quasicrystal and the pressurized approximant crystal, we have constructed the periodic Anderson model for the approximant crystal with 4f electrons at Yb sites 
and 3p electron at Al sites~\cite{WM2016}. 
Namely, we consider the system with periodic arrangement of the bcc structure of the Yb-Au-Al cluster (see Fig.~\ref{fig:YbAuAl}), which was discussed in section~2.2.
As a first step of analysis, we consider the case where Al/Au mixed sites framed in red in Figs.~\ref{fig:YbAuAl}(b) and \ref{fig:YbAuAl}(d) are occupied by Al and orbital degeneracy 
of electrons is neglected.  
The transfer integral of 3p electrons between the nearest-neighbor (N. N.) Al sites is set to be $t_2$ (see Fig.~\ref{fig:YbAuAl}(b)) and the f-c hybridization between 4f and 3p electrons at the N. N. Yb and Al sites on the 3rd shell (see Fig.~\ref{fig:YbAuAl}(c)) and 2nd shell (see Fig.~\ref{fig:YbAuAl}(b)) is set to be $V_0$. 
The transfer integral of 3p electrons between the N. N. Al sites on the 5th shell (see Fig.~\ref{fig:YbAuAl}(e)) is set to be $t_{5}'$. 
The other transfers and f-c hybridizations between the different shells are set so as to follow the distance dependence 
$1/r^{\ell+\ell'+1}$
with azimuthal quantum numbers, $\ell$, $\ell'$ (see Ref.~\citen{WM2016} for details)~\cite{Andersen1,Andersen2,Harrison}. 

\begin{figure}
\begin{center}
\includegraphics[width=15cm]{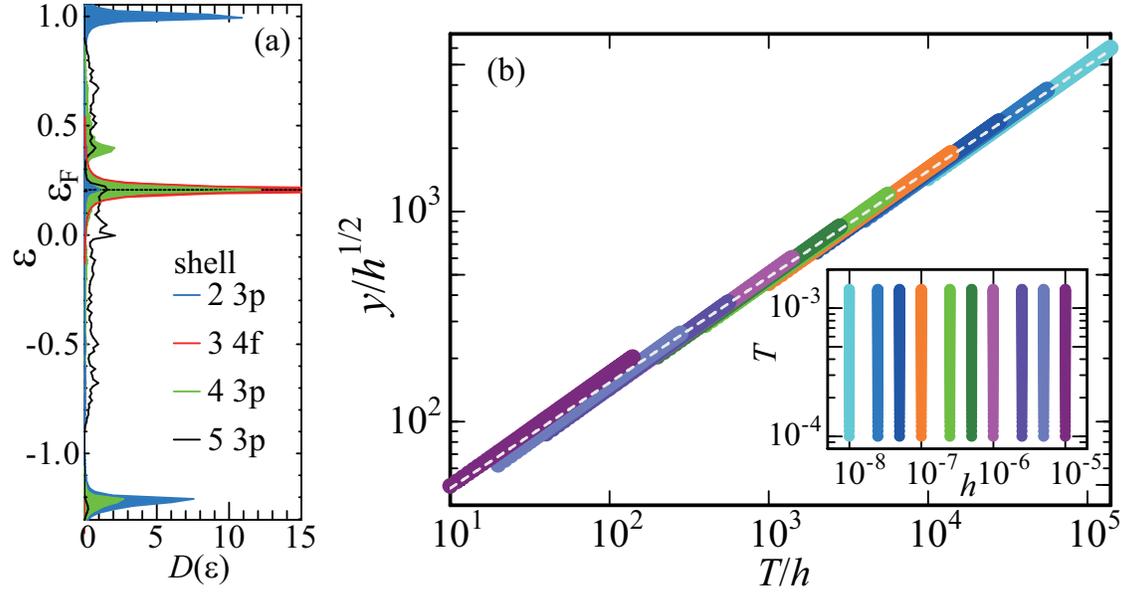}
\end{center}
\caption{(Color online) (a) Density of states of 4f electrons on the 3rd shell (red) and 3p electrons on the 2nd shell (blue), the 4th shell (green), and the 5th shell (black) in the Yb-Aul-Al cluster in the pressurized approximant crystal. The parameters are set as $t_2=1.0$, $t_5'=0.2$, $V_0=0.13$,  $\varepsilon_{\rm f}=-0.4$, and $U=\infty$ at half filling~\cite{WM2016}. Horizontal dashed line indicates the Fermi level $\varepsilon_{\rm F}$. (b) $T/h$ scaling at the QCP of valence transition at $U_{\rm fc}=0.0192$ for parameter set in (a)~\cite{WM2016}. Inset shows the data range in the $T$-$h$ phase diagram which scaling applies.}
\label{fig:TBscaling}
\end{figure}

By using the slave-boson mean-field theory for the strong limit of Coulomb repulsion between f electrons, $U=\infty$, for a typical parameter set of heavy electrons at half filling, 
the band structure is calculated~\cite{WM2016}.
In Fig.~\ref{fig:TBscaling}(a), the density of states of each 4f and 3p electrons on the 2nd to 4th shells, respectively, $D(\varepsilon)$, for the parameters,  
$t_2=1.0$, $t_{5}'=0.2$, $V_0=0.13$, $\varepsilon_{\rm f}=-0.4$, and $U=\infty$,   
is shown, which simulates the pressurized approximant crystal.   
Because of strong Coulomb repulsion between 4f electrons, the renormalized f level is raised up to the vicinity of the Fermi level $\varepsilon_{\rm F}$, giving rise to the heavy quasiparticle band~\cite{WM2016}. 
Hence, the dominant contribution to the density of states at the Fermi level comes from 4f electrons at Yb sites on the 3rd shell (see Fig.~\ref{fig:YbAuAl}(c)) and the next leading contribution comes from 3p electrons at Al sites on the 4th shell (see Fig.~\ref{fig:YbAuAl}(d)). This is due to the largest f-c hybridization reflecting the shortest Yb-Al distance (see Yb and Al sites indicated by arrows in Figs.~\ref{fig:YbAuAl}(c) and \ref{fig:YbAuAl}(d), respectively). 

Indeed, we found that the charge-transfer fluctuation between the 4f electron at Yb on the 3rd shell and the 3p electron at Al on the 4th shell is greatly enhanced~\cite{WM2016}. 
A remarkable result is almost dispersionless charge-transfer fluctuation mode appears around ${\bf q}={\bf 0}$, which is ascribed to strong local correlation effect for f electrons~\cite{WM2016}. 
To clarify how the local charge-transfer fluctuation affects the quantum criticality, let us focus on the charge-transfer mode. 
Then, we apply the recently-developed mode-mode coupling theory of CVF under magnetic field~\cite{WM2014} to the present system. Namely, starting from the periodic Anderson model with the inter-orbital Coulomb repulsion $U_{\rm fc}$ and the Zeeman term with magnetic field $h$, we have derived the self-consistent renormalization (SCR) equation for CVF~\cite{WM2016}. 
By solving the valence SCR equation at the QCP, which is identified to be $U_{\rm fc}=0.0192$ for the same parameter set used in Fig.~\ref{fig:TBscaling}(a), we found that the $T/h$ scaling 
\begin{eqnarray}
y=h^{1/2}\phi\left(\frac{T}{h}\right)
\label{eq:TBscaling}
\end{eqnarray}
appears over four decades in $T/h$, as shown in Fig.~\ref{fig:TBscaling}(b). 
The dashed line indicates the least-square fit of the data in the large-$T/h$ region, 
$y/h^{1/2}=c(T/h)^{\zeta}$, 
which gives the exponent $\zeta=0.5$. 
Then, the solution of the valence SCR equation turns out to have $y\sim T^{0.5}$ dependence. Hence, the magnetic susceptibility $\chi$ as well as the valence susceptibility $\chi_{\rm v}$ shows the criticality as $\chi\propto\chi_{\rm v}\propto y^{-1}\sim T^{-0.5}$. 
This result indicates that the magnetic susceptibility in the pressurized approximant crystal exhibits $\chi\sim T^{-0.5}$ for the zero-field limit and also the $T/H$ scaling behavior Eq.~(\ref{eq:TBscaling}) as observed in the quasicrystal~\cite{Deguchi} and also in the pressurized approximant crystal~\cite{Matsukawa2016}. 
Now, the characteristic temperature of CVF defined by Eq.~(\ref{eq:T0}) is evaluated to be $T_{0}\sim10^{-4}$ in the unit of $t_2$, which is comparable to the lowest temperature (see inset in Fig.~\ref{fig:TBscaling}(b)). 
By the detailed analysis of the mechanism of the $T/H$ scaling in $\beta$-YbAlB$_4$~\cite{WM2014},it turned out that the emergence of $T/H$ scaling is ascribed to the presence of the small characteristic temperature $T_0$.  

Then, let us discuss the quasicrystal from the viewpoint of infinite limit of the unit cell size of approximant crystals. 
Now, the small $T_0$ is realized by small coefficient $A$ due to strong locality of charge-transfer mode (see Eq.~(\ref{eq:chiv})) and the small Brillouin zone $q_{\rm B}$ reflecting the large unit cell. 
In reality, $3p_x$, $3p_y$, $3p_z$ energy bands of Al and 6s energy band of Au are considered to exist. Those conduction bands are folded into the small Brillouin zone, and hybridizations between them each other give rise to many splits into bonding and antibonding bands~\cite{Fujiwara}. 
Therefore, conduction bands themselves already have the flat nature. Hybridization between f and their conduction bands is expected to further promote locality of CVF. 
The quasicrystal corresponds to the infinite limit of the unit-cell size of approximant crystal. Then, the quasicrystal is regarded as the system with the small limit of the Brillouin zone. Hence, the characteristic temperature of CVF in the quasicrystal $T_0$ is considered to be at least smaller than the measured lowest temperature so that in the low temperature region, $T_{\rm F}^{*}\gg T\gsim T_{0}$, the quantum valence criticality is manifested in a series of physical quantities as 
$\chi\sim T^{-0.5}$, $(T_{1}T)^{-1}\sim T^{-0.5}$, $\rho\sim T$, and $C/T\sim -\log{T}$ (see Table~\ref{tb:QCP}). 

\section{Experimental indication of valence instability}

To clarify the orign of the unconventional quantum criticality listed in Table~\ref{tb:QCP} definitely, experimental identification of the QCP of the valence transition is highly desired.
In YbCu$_{5-x}$Al$_{x}$, it has been observed that the Yb-valence crossover occurs near $x=1.5$~\cite{Bauer1997}. 

To detect the evidence of the valence QCP, materials have been explored by performing the   measurements of Yb valence and magnetic susceptibility in extreme condition of low temperature, high pressure, and strong magnetic field. 
Recently, the evidence of the CEP of the first-order valence transition has been discovered in YbNi$_3$Ga$_9$ by Matsubayashi and coworkers~\cite{Matsubayashi}. 

In YbRh$_2$Si$_2$, careful measurement by NMR at low temperatures under magnetic field has been performed by Kambe, in which a signature of phase separation of two heavy-electron states was reported~\cite{Kambe2014}. This suggests that relatively larger- and smaller-Yb-valence states coexist, indicating close proximity to the first-order Yb-valence transition.    

Consequences of these experimental indications of valence instability will be discussed in the following subsections. 

\subsection{Discovery of CEP of the first-order valence transition in YbNi$_3$Ga$_9$}

By constructing the $T$-$P$-$H$ phase diagram of YbNi$_3$Ga$_9$,  the CEP of the first-order valence transition has been discovered in the paramagnetic-metal phase~\cite{Matsubayashi}. It is also observed that the magnetic susceptibility is enhanced toward the CEP, which detects the signature of simultaneous divergence of the valence susceptibility and the uniform magnetic susceptibility at the CEP predicted by the CVF theory~\cite{WM2008,WTMF2009,WM2010}. 
It is found that the first-order valence transition surface extends to the magnetically-ordered phase for the $P>P_{\rm c}\approx 9$~GPa region, giving rise to the first-order magnetic transition surface. Namely, the first-order valence transition is considered to terminate the magnetic order, which has been theoretically shown to occur in the ${\cal H}_{\rm EPAM}$~\cite{WM_CeRhIn5,WM2012}. 
Interestingly, the ``critical end line" which connects CEP's of the valence transition $T_{\rm  CEP}(P,H)$ in the $T$-$P$-$H$ phase diagram~\cite{Matsubayashi} seems to be continued to the ``tricritical line" which connects the tricritical points of the magnetically-ordered phase $T_{\rm TCP}(P,H)$ (see Fig.4 in Ref.~\citen{Matsubayashi}). 

It is also interesting to identify the location of the quantum CEP defined by $T_{\rm  CEP}(P_{\rm QCP},H_{\rm QCP})$=0,  i.e., the QCP of the valence transition, in the $T$-$P$-$H$ phase diagram. 
By tuning the pressure and magnetic field to $P_{\rm QCP}$ and $H_{\rm QCP}$, respectively, temperature dependence of physical quantities such as 
$\rho(T)$, $C(T)/T$, $\chi(T)$, and $(T_1T)^{-1}$ can be observed. At the QCP, the criticality of the CVF theory listed in Table~\ref{tb:QCP} is expected to appear. 
Such measurements are highly desired, which are left for future studies.

\subsection{Signature of coexistence of two-component heavy electron states in YbRh$_2$Si$_2$}

Recently, in YbRh$_2$Si$_2$, NMR measurement has revealed that two heavy-electron states coexist at low temperatures under magnetic field~\cite{Kambe2014}. This suggests that phase separation of higher and lower valences of Yb occurs, which can be  naturally understood if this material is located closely to the first-order valence transition. 
Actually, it has been shown theoretically that the CEP of the first-order valence transition can be induced by applying magnetic field in the ${\cal H}_{\rm EPAM}$~\cite{WM2008,WTMF2009}. 
This implies that the CEP is raised up to approach 0~K (see Fig.~\ref{fig:T_P}(d)) and eventually appears at the $T>0$ side (see Fig.~\ref{fig:T_P}(b)) by applying magnetic field, even if the CEP is located at the $T<0$ side for zero magnetic field. 
In reality, the first-order valence transition line is expected to be surrounded by the phase-separated region of relatively-higher and -lower valence states of Yb due to extrinsic effects such as 
lattice defect and impurity incorporation, which are inevitable in real materials. 
The experimental fact that the coexistence region expands in the $T$-$H$ phase diagram as $H$ increases~\cite{Kambe2014} is in accord with this picture.   

\section{Summary}

The theory of CVF is shown to exhibit a new universality class of quantum criticality. 
Quantum valence criticality gives a unified explanation for the unconventional criticality 
observed in Yb-based periodic crystals and also the ``robustness" of the criticality observed 
in the Yb-based quasicrystal under pressure. 
The CVF theory has shown  
the emergence of the common quantum criticality even in the approximant crystal by pressure tuning and 
the wider quantum critical region in the quasicrystal than in the approximant crystal in the $T$-$P$ phase diagrams. These predictions have been actually observed by recent experiments.
It also predicts that 
the valence-crossover region appears in the quasicrystal under magnetic field.

By constructing the periodic Anderson model on the approximant crystal, we have recently found that charge-transfer fluctuation between 4f electron at Yb on the 3rd shell and 3p electron at Al on the 4th shell in the Yb-Au-Al cluster is dominant, which has strong local character. 
By applying the mode-mode coupling theory to this mode, we have shown that the small characteristic temperature $T_0$ of the CVF emerges so that the quantum valence criticality and the $T/H$ scaling  behavior appear. 
This gives a unified explanation for the common criticality in the approximant crystal under pressure and the quasicrystal. 
These results suggest that a new class of universality is formed irrespective of periodic crystal and quasicrystal because of strong locality of CVF.

As an experimental indication of valence instability, 
evidence of the critical point of the first-order valence transition has been recently observed 
in the $T$-$P$-$H$ phase diagram of YbNi$_3$Ga$_9$. 

High-pressure measurements have been playing key roles for revealing these new aspects of quantum critical phenomena, which are also expected to play significant roles for future studies toward complete understanding of the new quantum critical phenomena.

\acknowledgment

One of the authors (S. W.) acknowledges SPring-8 long-term proposal 
``Experimental verification of new quantum critical phenomena of critical valence fluctuations 
by X-ray spectroscopy" (Proposal No.~0046 in 2012B, 2013A, 2013B, 2014A, 2014B, and 2015A). 
The experimental result of YbNi$_3$Ga$_9$ was obtained in collaboration with 
K.~Matsubayashi, T. Hirayama, T.~Yamashita, S. Ohara, N. Kawamura, M. Mizumaki, N.~Ishimatsu, K. Kitagawa, and Y. Uwatoko.
The authors thank S. Matsukawa, K. Deguchi, N. K. Sato, and T. Ishimasa for showing us their experimental data with enlightening discussions. 
Theoretical works were supported by JSPS KAKENHI Grant Numbers JP24540378, JP25400369, 
JP15K05177 and JP16H01077A01 from Japan Society for the Promotion of Science (JSPS).

\end{document}